\documentclass[reprint,
superscriptaddress,
longbibliography,
amsmath,
amssymb,
aps
]{revtex4-2}

\usepackage{epsfig,amsmath,amssymb,bm,epsf,psfrag,bbm}
\usepackage{graphicx}
\usepackage{dcolumn}
\usepackage{bm}
\usepackage{amsbsy}
\usepackage{amsthm}
\usepackage{color, comment, verbatim}
\usepackage{ragged2e}
\usepackage{subfigure}
\usepackage{enumerate}
\usepackage{chngcntr}
\usepackage[T1]{fontenc}
\usepackage[english]{babel}
\usepackage[utf8]{inputenc}
\usepackage{dsfont} 
\usepackage{tikz}
\usepackage[toc,page]{appendix}
\usepackage[colorlinks,breaklinks]{hyperref} 
\usepackage{xcolor}
\usepackage{titlesec}

\makeatletter
\newcommand\org@hypertarget{}
\let\org@hypertarget\hypertarget
\renewcommand\hypertarget[2]{%
  \Hy@raisedlink{\org@hypertarget{#1}{}}#2%
  }
\makeatother

\hypersetup{
	bookmarksnumbered,
	pdfstartview={FitH},
	citecolor={darkgreen},
	linkcolor={darkred},
	urlcolor={darkblue},
	pdfpagemode={UseOutlines}}
\definecolor{darkgreen}{RGB}{0,0,0}
\definecolor{darkblue}{RGB}{0,0,0}
\definecolor{darkred}{RGB}{0,0,0}

\usepackage{booktabs}
\usepackage{tablefootnote}
\usepackage{threeparttable}
\usepackage{color,soul}

\usepackage{natbib} 
\setcitestyle{sectionbib}
\usepackage{bibunits}
\defaultbibliographystyle{apsrev4-1fixed_with_article_titles_full_names}
\defaultbibliography{reference_list.bib}

\newcommand{\figsisimul}{S1}
\newcommand{\figsiunitary}{S2}
\newcommand{\figsilocal}{S3}
\newcommand{\figsiaxial}{S4}
\newcommand{\figsisetup}{S5}

\title{
Three-dimensional holographic imaging of incoherent objects through scattering media}
\makeatletter\let\Title\@title\makeatother

\begin{document}

\title{\Title}

\author{YoonSeok Baek}
 \email{yoonseok.baek@lkb.ens.fr}
    \affiliation{Laboratoire Kastler Brossel, ENS–Université PSL, CNRS, Sorbonne Université, Collège de France, 24 Rue Lhomond, F-75005 Paris, France}
    
\author{Hilton B. de Aguiar}
    \affiliation{Laboratoire Kastler Brossel, ENS–Université PSL, CNRS, Sorbonne Université, Collège de France, 24 Rue Lhomond, F-75005 Paris, France}
    
\author{Sylvain Gigan}
    \affiliation{Laboratoire Kastler Brossel, ENS–Université PSL, CNRS, Sorbonne Université, Collège de France, 24 Rue Lhomond, F-75005 Paris, France}
    
\begin{bibunit}
\begin{abstract}
Three-dimensional (3D) high-resolution imaging is essential in microscopy, yet light scattering poses significant challenges in achieving it. Here, we present an approach to holographic imaging of spatially incoherent objects through scattering media, utilizing a virtual medium that replicates the scattering effects of the actual medium. This medium is constructed by retrieving mutually incoherent fields from the object, and exploiting the spatial correlations between them. By numerically propagating the incoherent fields through the virtual medium, we non-invasively compensate for scattering, achieving accurate 3D reconstructions of hidden objects. Experimental validation with fluorescent and synthetic incoherent objects confirms the effectiveness of this approach, opening new possibilities for advanced 3D high-resolution microscopy in scattering environments.
\end{abstract}

\maketitle

Visualizing the 3D structures of materials and biological specimens is fundamental to various scientific fields.
To capture these structures at the microscopic scale, conventional microscopy techniques achieve 3D imaging by controlling excitation beams \cite{mertz2019strategies}, employing engineered point spread functions \cite{levoy2006light,pavani2009three,liu2017multiplexed}, or utilizing holographic approaches \cite{schilling1997three,rosen2008non,yu2014review,memmolo2015recent}. However, these methods, which depends on precisely controlled excitation and detection schemes, face significant limitations in highly scattering environments where light scattering generates random speckle patterns, obscuring spatial information.

Over the past decade, several methods have been developed for non-invasive imaging through scattering media \cite{yoon2020deep,bertolotti2022imaging}. 
For imaging incoherent objects, such as those composed of fluorescent sources, wavefront shaping techniques achieve focused excitation through scattering media by exploiting nonlinear fluorescence signals \cite{katz2014noninvasive,papadopoulos2017scattering,may2021fast,zhao2024single} or optimizing specific metrics \cite{boniface2019noninvasive,daniel2019light,aizik2022fluorescent,aizik2024non}. 
Alternatively, computational imaging using scattered light \cite{bertolotti2012non,katz2014non,boniface2020non,zhu2022large,weinberg2023noninvasive,wu2017imaging} has been demonstrated by exploiting speckle correlations \cite{freund1988memory}. 
Despite their differences, both approaches are generally restricted to two-dimensional (2D) imaging of small areas, due to the limited range of focus scanning and speckle correlations.
More recently, efforts have been made towards 3D incoherent imaging through scattering media \cite{matsuda20243d1,matsuda20243d2,okamoto2019noninvasive,horisaki2019single,shi2017non,aarav2024depth}, but they rely on simplifying assumptions such as treating fluorescence as fully coherent field or requiring the object to be distant from the scattering plane. 

Here, we introduce a method for holographic imaging of spatially incoherent objects obscured by scattering media. Our approach begins by retrieving multiple incoherent fields through spatial modulation of scattered light, as recently demonstrated in \cite{baek2023phase}. Leveraging the inherent correlation within the scattered fields, we extract wavefront distortions at a specific internal plane to build a virtual medium that replicates the scattering effects of the actual medium. By numerically propagating the scattered fields through this virtual medium, our method effectively compensates for the scattering, reconstructing 3D images of the hidden objects. This field-based approach extends non-invasive imaging capabilities beyond the conventional limits of intensity-based methods, extending both lateral and axial imaging ranges. 
We experimentally demonstrate the effectiveness of this method by imaging incoherent object, composed of fluorescent and synthetic sources, through scattering layers.

\section*{Results}
Our method constructs a virtual medium that replicates the scattering behavior of the actual physical medium by leveraging the correlation between scattered fields. 
\begin{figure*}[ht]
	\centering
	\includegraphics[width=1.9\columnwidth]{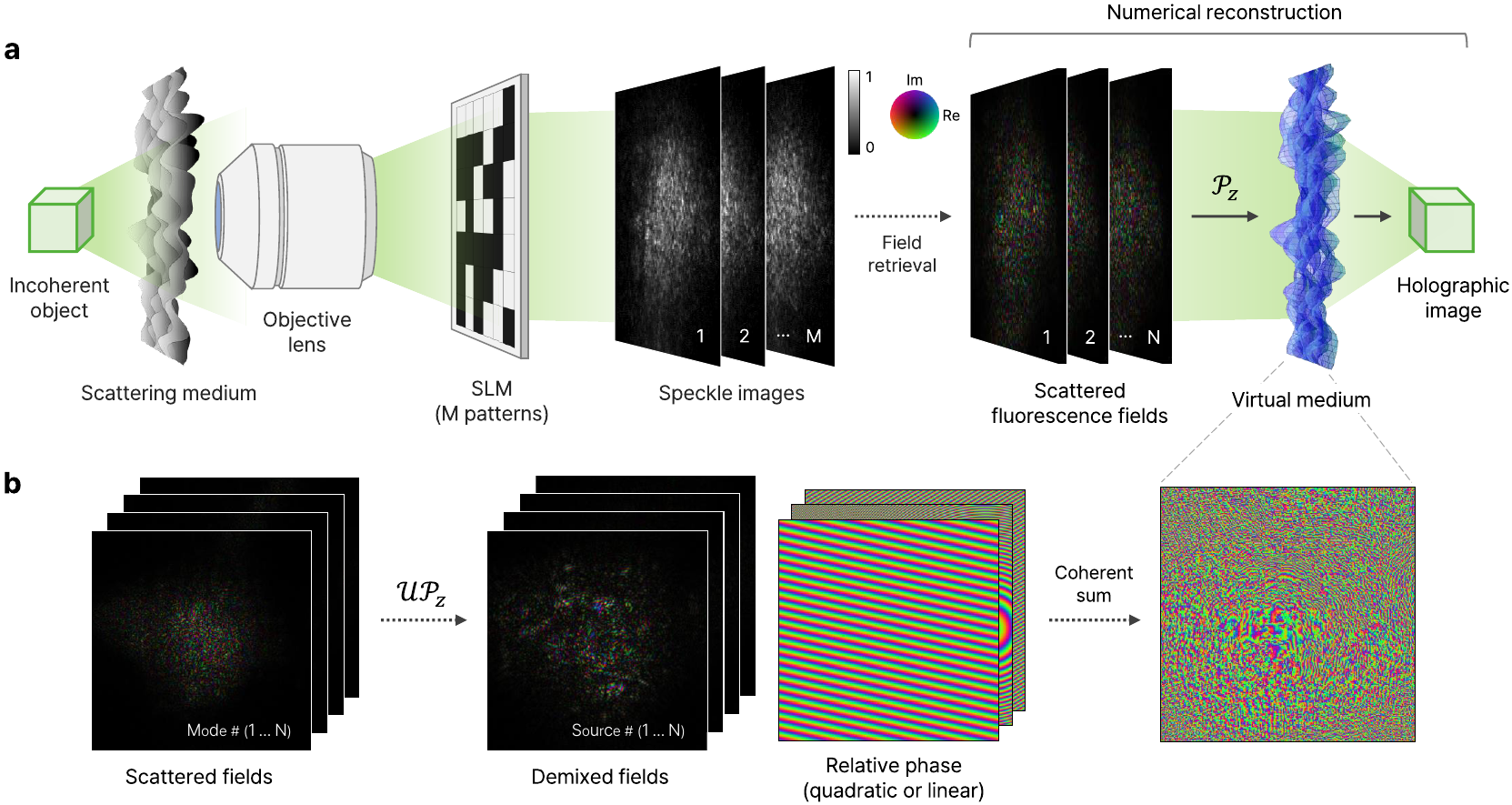}
	\caption{\textbf{Holographic reconstruction of incoherent objects using virtual medium}: \textbf{a}, Experimental setup and image reconstruction. The scattering medium scatters the incoherent light emitted by an object. The scattered light is then spatially modulated by random patterns on a spatial light modulator (SLM). The resulting incoherent speckle images are used to retrieve scattered fields using the phase retrieval algorithm. By numerically propagating the fields through a virtual scattering medium, the image of the object is obtained. \textbf{b}, Reconstruction of the virtual medium. The retrieved fields are numerically propagated ($\mathcal{P}_z$) and unitary transformed ($\mathcal{U}$) into the scattered fields for individual sources. After correcting their relative phase (quadratic or linear), the scattered fields are coherently summed to construct the virtual medium.}
	\label{fig.1}
\end{figure*}
To measured the fields of spatially incoherent light, we employed a phase retrieval method that we recently introduced in \cite{baek2023phase}. This method spatially modulates the incoherent light using random phase masks. Then the resulting speckle pattern is analyzed through an iterative algorithm to retrieve mutually incoherent fields of independent incoherent sources (see Methods). 

The underlying correlation between these scattered fields generally manifests as shifts and tilts in their wavefronts, for neighboring, laterally shifted sources \cite{osnabrugge2017generalized}. This correlation can be expressed as
\begin{equation}
C(\Delta r, \Delta k)=\int E_n(r-\Delta r)E_m^*(r)e^{i\Delta k \cdot r}d^2r.
\end{equation}
Here $E_n(r)$ is the scattered field from an $n$-th source, corresponding to a column of an outgoing transmission matrix. In forward-scattering scenarios, this correlation reaches its maximum when the shift $\Delta r$ and tilt $\Delta k$ are optimally adjusted \cite{osnabrugge2017generalized}. At a specific internal plane, these adjustments are equivalent to a wavefront tilt without any lateral shift:
\begin{equation}
\max C = \int \mathcal{E}_n(r,r_n)\mathcal{E}_m^*(r)e^{i\Delta k_\textnormal{opt} \cdot r}d^2r,
\label{eq.C_max}
\end{equation} 
where $\mathcal{E}_n(r) = P_{\hat{z}}\left[{E_n(r)}\right]$, $P_{\hat{z}}$ is the propagation to the plane where the tilt correlation is maximized, and $\Delta k_\textnormal{opt}$ is the optimal wavefront tilt. This expression implies that, after accounting for the wavefront tilt, scattered fields retain a consistent wavefront distortion caused by scattering.
We model this distortion as a virtual scattering layer $S(r)$, at the internal plane, such that $\mathcal{E}_n(r) \approx S(r)h_n(r)$, where $h_n(r)$ represents the field emitted by the source under free-space propagation (See Methods for locating the correlation plane). 
Although the argument above primarily applies to laterally shifted sources, it can be extended to three-dimensional objects. Specifically, if the axial extent of the incoherent object is not significantly larger than the scattering mean free path, the phase of $S(r)$ can be expressed as
\begin{equation}
\text{arg}[S(r)] = \text{arg}\left[\sum_n \mathcal{E}_n(r)e^{i\Phi_{nm}(r)}\right],
\label{eq.S}
\end{equation}
where $\Phi_{nm}(r)$ represents the quadratic phase term, given by the phase of $h_n^*(r)h_m(r)\approx \mathcal{E}_n^*(r)\mathcal{E}_m(r)$, and $m$ is an arbitrary index. 
For 2D objects, $\Phi_{nm}(r)$ reduces to the optimal wavefront tilt in Eq. \ref{eq.C_max}. The virtual scattering layer in Eq. \ref{eq.S} captures the wavefront distortion that is common to all scattered fields relative to $\mathcal{E}_m^*(r)$. By propagating the fields through this layer, the scattering effects can be effectively compensated, resulting in a 3D image of the incoherent object. The holographic image is then obtained as: 
\begin{equation}
    I(r,z) = \sum_{n} \left| \mathcal{P}_z \left\{ 
    \mathcal{F} \left[ 
    \mathcal{E}_n^*S \right](r) 
    \right\} \right|^2,
    \label{eq.3D_intensity}
\end{equation}
where $\mathcal{F}$ is the Fourier transformation (see Supplementary for further details). 
In practice, accurately determining the relative phase $\Phi_{nm}(r)$ could be challenging when there is the limited spatial overlap between the fields $\mathcal{E}_n(r)$ and $\mathcal{E}_m(r)$. To address this, we initially constructed $S(r)$ using a pair of fields with high correlation, then incrementally update $S(r)$ by incorporating additional fields that show strong correlation (see Methods).

\begin{figure*}[ht]
	\centering
	\includegraphics[width=1.9\columnwidth]{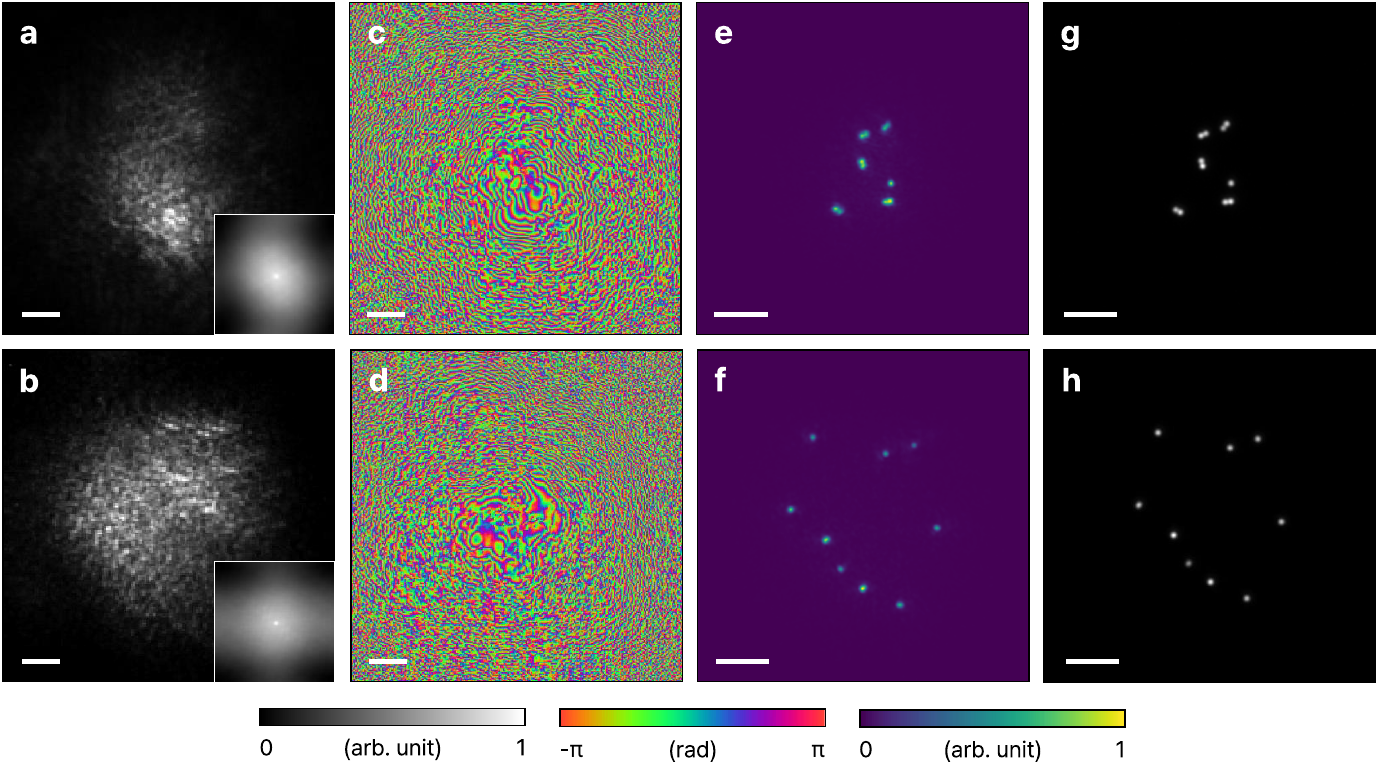}
	\caption{\textbf{Imaging 2D fluorescent objects through separate scattering layers
}: \textbf{a--b}, Fluorescence speckle measured at the camera. (Inset) Background-subtracted speckle auto-correlation. \textbf{c--d}, Virtual scattering layers constructed using the scattered fluorescence fields. \textbf{e--f}, Image obtained by numerically propagating the fields through the virtual scattering layers. \textbf{g--h}, Ground truth fluorescence images measured from the side without the scattering layers. Scale bars represent 10 $\mu$m.
}
	\label{fig.2}
\end{figure*}

To experimentally demonstrate the method, we used fluorescent beads placed approximately 100 $\mu m$ behind a highly scattering layer (see Methods for experimental setup). The scattered fluorescence produced an incoherent sum of speckle patterns (Fig. 2a, b), and the speckle autocorrelation failed to reveal information about the object (Fig. 2a, b insets). This is because, under microscopic imaging conditions with the object near the scattering layer, scattering cannot be approximated as a convolution of the object with a single speckle pattern \cite{hofer2018wide}. Despite this limitation, our method successfully retrieved incoherent scattered fields and constructed the virtual scattering layer (Fig. 2c, d). Each virtual layer was numerically positioned at the plane where the field correlation was highest. Finally, images of the hidden fluorescent objects were reconstructed by numerically propagating the fields through the virtual layer and summing their intensities based on Eq. \ref{eq.3D_intensity} (Fig. 2e, f). The reconstructed images are in strong agreement with the fluorescence images taken without the scattering layer (Fig. 2g, h), which confirms the effectiveness of our method.

An important advantage of our holographic approach is its capability for 3D imaging through scattering media. To demonstrate this, we placed fluorescent beads at two different depths behind the scattering layer (see Methods). As in previous experiments, the scattered fluorescence produced incoherent speckle patterns on the camera (Fig. 3a). We then constructed a virtual scattering layer and numerically propagated the scattered fields through it (see Fig. 3b and Methods). Figure 3c shows the resulting 3D intensity distribution from this propagation, where $z$ represents the distance from the physical scattering layer. The magnification and z-coordinate of the image were adjusted for accurate representation (see Supplementary). The result clearly reveals the 3D distribution of fluorescent sources, with images at the two planes highly consistent with those obtained without the scattering layer (Fig. 3d-g).
\begin{figure*}[ht]
	\centering
	\includegraphics[width=1.9\columnwidth]{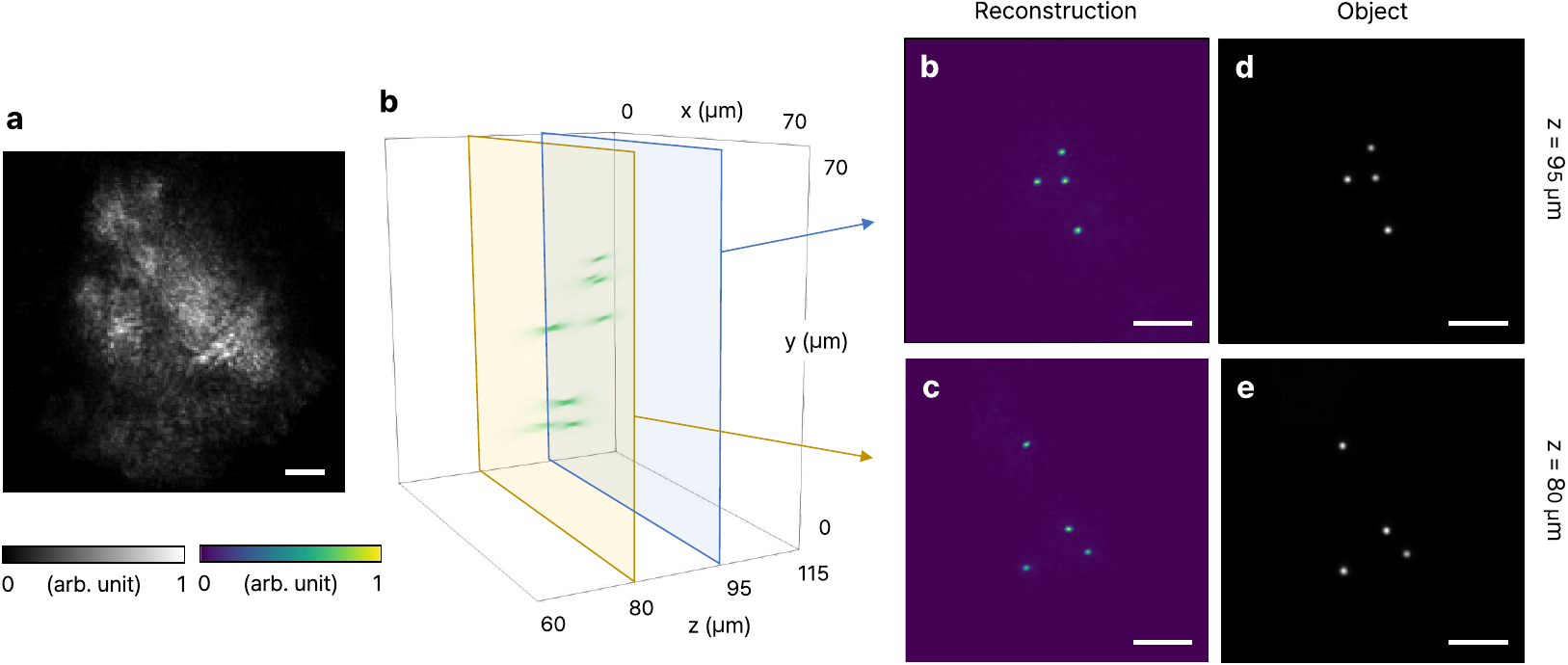}
	\caption{\textbf{Imaging a 3D fluorescent object through a scattering layer
}: \textbf{a}, Fluorescence speckle measured at the camera. \textbf{b}, Virtual scattering layers constructed using the scattered fluorescence fields. \textbf{c}, 3D rendering of the numerical propagation. \textbf{d--e}, Image reconstructed at 80 and 95 um away from the virtual layer. \textbf{f--g}, ground truth image obtained from the side without the scattering layer. Scale bars represent 10 $\mu$m.
}
	\label{fig.3}
\end{figure*}

\begin{figure*}[ht]
	\centering
	\includegraphics[width=2\columnwidth]{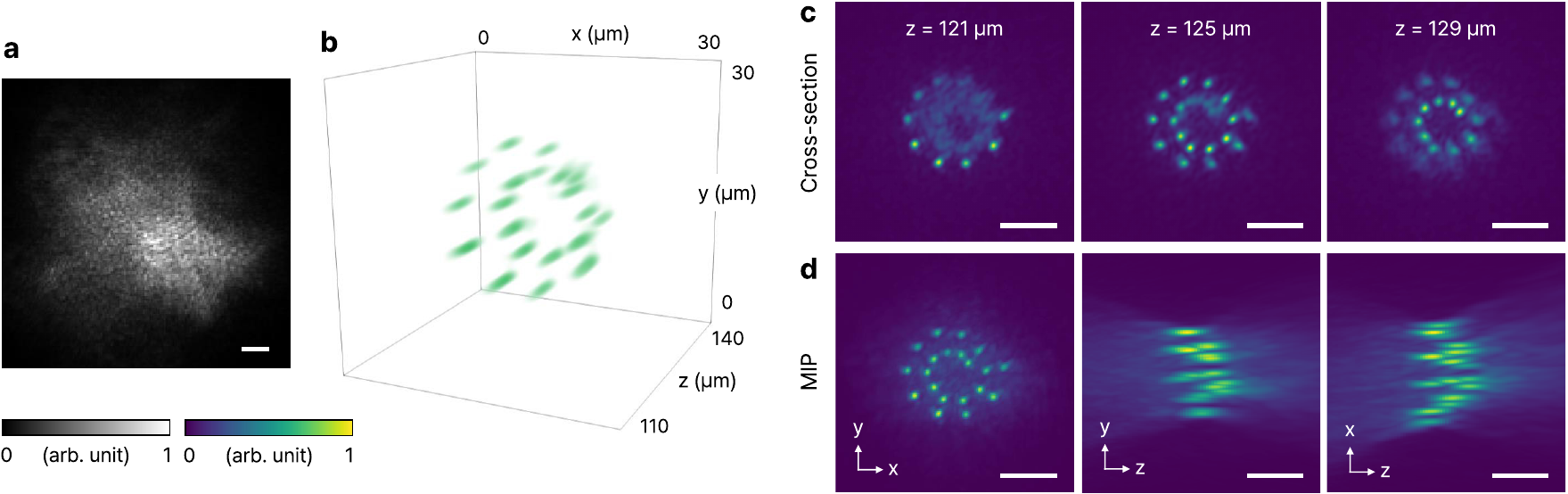}
	\caption{\textbf{Imaging a complex synthetic incoherent object
through a scattering layer}: \textbf{a}, Experimental speckle image of the synthetic incoherent object arranged in a spiral. \textbf{b}, Rendered 3D reconstruction of the object. \textbf{c}, Cross-sectional view of the reconstruction. \textbf{c}, Maximum intensity projection (MIP) images of the reconstruction. Scale bars represent 10 $\mu$m.
}
	\label{fig.4}
\end{figure*}

To demonstrate the capability of our method in handling more complex scenarios, we tested its performance with sources distributed across multiple axial planes using a synthetic 3D incoherent object. 
The incoherent data were generated by projecting diffraction-limited laser foci, corresponding to point sources, in a 3D volume using a digital micromirror device (see Methods), which served as ground truth. Speckle images from all point sources were measured and summed to simulate the final incoherent image (Fig. \ref{fig.4}a), which was then processed by the algorithm as if it originated from a single incoherent object.
In this experiment, 20 diffraction-limited foci were arranged along a spiral trajectory described by the parametric equation  $(r_n,\theta_n) = (r_1-r_2\frac{n-1}{20}, 4\pi\frac{n-1}{20})$, where $n$ is the index of each source, $r_1 = 15 \mu m$ and $r_2 = 5 \mu m$. The axial position of each source was increased by 0.5 $\mu m$, resulting in a total span of 10 $\mu m$. 
The reconstructed 3D image successfully captured the spatial distribution of the incoherent object (Fig. \ref{fig.4}b). The 3D positions of the sources were clearly visualized in cross-sectional views (Fig. \ref{fig.4}c) and maximum intensity projections (MIPs, Fig. \ref{fig.4}d). To further demonstrate the versatility of the method, we conducted numerical simulations with incoherent objects shaped as three perpendicular lines and cube edges (Fig. \figsisimul). These objects were then obscured by a highly scattering layer that uniformly scatters light with a 90-degree divergence (Fig. \figsisimul b). Using our method, we successfully reconstructed the 3D structures of these object (Fig. \figsisimul c). These results highlights the robustness of the method in imaging complex 3D continuous objects.

\section*{Discussion}
We have demonstrated holographic imaging of spatially incoherent objects through scattering media. Our approach leverages mutually incoherent scattered fields to construct a virtual scattering medium. By numerically propagating the scattered fields through the virtual medium, we effectively compensate for the scattering effects and reconstruct 3D images of the hidden objects. The virtual medium acts as a planar scatterer at an internal plane, capturing the common distortions shared by light originating from different sources. This suggests that extending the method to volumetric media is feasible, as such common distortions in biological tissue have been experimentally demonstrated \cite{badon2020distortion}.

A key aspect of our method is its use of the inherent field correlation in scattered light \cite{osnabrugge2017generalized}. This enables non-invasive 3D imaging in a wider range of scenarios, unlike axial memory effect-based methods, which require the object to be several millimeters away from the scattering plane and have an axial extent much smaller than this distance. \cite{hofer2018wide,may1977information,aarav2024depth,shi2017non,horisaki2019single,okamoto2019noninvasive}.
Maintaining stability is an important experimental factor because the setup instability can reduce the speckle contrast more than is expected from the low-spatial coherence of the object, making accurate retrieval of scattered fields difficult. This could be mitigated by employing a position refinement algorithm and short exposure measurements \cite{aidukas2024high}.
While our method does not actively control excitation, it could also benefit from dynamic excitation schemes utilized in recent studies \cite{boniface2019noninvasive,zhu2022large,weinberg2023noninvasive,lee2022high}.
In conclusion, our work paves the way for 3D imaging in highly scattering environments. We anticipate that this method will open up new possibilities for imaging and sensing applications involving the scattering of spatially incoherent light.

\section*{Methods}
\subsection*{Demixing fields for individual sources}
Mutually incoherent fields and their unitary transformations are indistinguishable by intensity. 
As a result, the spatially incoherent fields are retrieved as linear mixtures of the scattered fields from individual sources \cite{baek2023phase}. To recover the fields corresponding to each sources, we utilize the difference in correlation between mixed and demixed fields. As described in the main text, the relative phase between fields from individual sources at the correlation plane can be approximated by a quadratic phase, $\mathcal{E}_n^*(r)\mathcal{E}_m(r)\approx e^{i\Phi_{nm}(r)}$. 
This indicates that the relative phase of the mixed fields is expressed by multiple quadratic phase terms. In contrast, the demixed fields have a relative phase with a single quadratic phase. 
Using this property, an appropriate unitary matrix $U$ for demixing can be identified. For example, in the case of 2D objects, the relative phase $\Phi_{nm}(r)$ reduces to a linear term. Based on this observation, we define a demixing metric as:
\begin{equation}
    M(U)=\sum_{n\neq m}{\max_{r\neq0} \left|\mathcal{F}\left[\mathcal{E}^{'*}_n\mathcal{E}^{'}_m\right](r)\right|},
\end{equation}
where $\mathcal{E}^{'}_n(r)= \sum_{m=1}^{N}U_{nm}\mathcal{E}^{\textnormal{ret}}_m(r)$, and $\mathcal{E}^{\textnormal{ret}}_n(r)$ is retrieved field at the correlation plane, defined as $\mathcal{E}^{\textnormal{ret}}_n(r)=P_{\hat{z}}\left[{E^{\textnormal{ret}}_n(r)}\right]$. 
This metric is maximized when the relative phase of the unitary-transformed fields $\mathcal{E}^{'}(r)$ exhibit a single linear term.
We determine the unitary matrix $U$ that maximizes $M$ using Riemannian optimization algorithm \cite{abrudan2008steepest}. 
The resultant unitary matrix effectively demixes the fields, $\mathcal{E}_n(r)= \sum_{m=1}^{N}U_{nm}\mathcal{E}^{ret}_n(r)$ (see Fig. \figsiunitary).

For 3D objects, the demixing metric for the 2D case can be extended to accommodate the quadratic phase of $\Phi_{nm}(r)$. However, rather than directly incorporating the quadratic phase, we employed a simplified approach that leverages local field correlations. By examining field correlations within small areas at the correlation plane, the quadratic phase of $\Phi_{nm}(r)$ can be locally approximated as a linear phase, making demixing possible similar to the 2D case. Figure {\figsilocal} illustrates the differences in field correlation (Eq. \ref{eq.wavefront_corr}) using the entire field of view and smaller regions. Each locally demixed fields are used to construct a virtual layer at their corresponding regions, and the entire virtual layer formed by stitching these local constructions. For example, the result in Fig. \ref{fig.4} was obtained by dividing the field of view into a 3$\times$3 grid of partially overlapping regions . 

\subsection*{Locating the correlation plane}
To determine the position of the internal correlation plane, we propagate the retrieved fields, $E^{\textnormal{ret}}_n$, and evaluate their tilt-correlation at different depths. To achieve this, we used the correlation function defined as:
\begin{equation}
   \Gamma(r,z)=\sum_{n,m}{\left|\mathcal{F}
   \left\{\mathcal{P}_z[E^{\textnormal{ret}}_n]^*\mathcal{P}_z[E^{\textnormal{ret}}_m]\right\}(r)
   \right|^2}.
   \label{eq.wavefront_corr}
\end{equation}
Then the axial position of the correlation plane, $\hat{z}$, is identified as the plane with the maximum correlation value:
\begin{equation}
\hat{z} = \arg\max_{z}\left[{\max_{r\neq0}{\Gamma(r,z)}}\right],
\end{equation}
where the contribution of autocorrelation is eliminated by excluding $r = 0$ (see Fig. \figsiaxial).

\subsection*{Experimental setup}
The experimental setup is illustrated in Fig. \figsisetup. For the results shown in Figs. \ref{fig.2}--\ref{fig.3}, a laser (Compass 215M-50, Coherent) was used to excite fluorescent beads (1 µm, F8820, Invitrogen), which were immersed in glycerol for the experiments in Fig. 2 and in UV glue (NOA 68, Norland) for those in Fig. 3. The emitted fluorescence was scattered by a 220-grit ground glass diffuser and collected by an objective lens (MPlan N 50× 0.75, Olympus). Scattered fluorescence was modulated by a spatial light modulator (SLM, X10468-04, Hamamatsu) and recorded by a sCMOS camera (PCO.edge 5.5, PCO) located at the Fourier plane of the SLM. To measure the emission signal while blocking the excitation beam, fluorescence filters (NF533-17 and MF530-43, Thorlabs) were placed in front of the sCMOS camera, with an additional filter (BLP01-532R-25, Semrock) positioned before the SLM. Ground truth images were obtained by directly imaging the fluorescent objects from the opposite side without the scattering media, using a different objective lens (Plan N 20× 0.4, Olympus) and a camera (acA5472-17um, Basler) with a fluorescence filter (NF533-17, Thorlabs). For the results in Fig. \ref{fig.4}, a digital micromirror device (AJD-4500, Ajile) was installed in the laser path (dashed box in Fig. \figsisetup), and the axial position of the tube lens (L1 in Fig. \figsisetup) was adjusted to project point sources at different depths. A bandpass filter (FL532-3, Thorlabs) was placed in front of the sCMOS camera, replacing the previously mentioned filters.

\section*{Acknowledgement}
This research was supported by European Union’s Horizon Europe research and innovation programme (MSCA-IF  N°101105899), H2020 Future and Emerging Technologies (863203), and European Research Council (724473), S.G. is member of the Institut Universitaire de France

\putbib
\end{bibunit}

\end{document}